\documentstyle[11pt,aaspp]{article}
\input epsf
\begin{document}
\title{The Shapes of Dense Cores and Bok Globules} 
\author{ Barbara S. Ryden \altaffilmark{1}}
\affil{ Department of Astronomy, The Ohio State University, \\
174 W. 18th Ave., Columbus, OH 43210}
\altaffiltext{1}{National Science Foundation Young Investigator;
ryden@mps.ohio-state.edu}

\begin{abstract}
The shapes of isolated Bok globules and embedded dense cores
of molecular clouds are analyzed using a nonparametric
kernel method, using the alternate hypotheses that they
are randomly oriented prolate objects or that they are
randomly oriented oblate objects. In all cases, the prolate
hypothesis gives a better fit to the data.
If Bok globules are oblate spheroids, they must be very flattened;
the average axis ratio is $< \! \gamma \! > \approx 0.3$, and
no globules can have $\gamma \gtrsim 0.7$. If Bok globules are
prolate, their intrinsic flattening is not as great, with
a mean axis ratio $< \! \gamma \! > \approx 0.5$.
For most data samples of dense cores embedded within molecular
clouds, the randomly-oriented oblate hypothesis can be rejected at the $99\%$
one-sided confidence level. If the dense cores are prolate, their
mean axis ratio is in the range $< \! \gamma \! > = 0.4 \to 0.5$.
Analysis of the data of Nozawa et al. (1991) reveals that dense
cores are significantly different in shape from the clouds in
which they are embedded. The shapes of dense cores are
consistent with their being moderately flattened prolate
spheroids; clouds have flatter apparent shapes, and are
statistically inconsistent with a population of axisymmetric
objects viewed at random angles.

\end{abstract}


\section{Introduction}
The molecular gas in our galaxy shows structure on a wide range of
scales. The largest structures detected in the molecular gas are
giant molecular clouds (GMCs), of which the largest have diameters
of $\sim 100 \, {\rm pc}$ and masses of more than $10^6 \, {\rm M}_{\sun}$.
High resolution studies of molecular clouds, however, reveal that
they have internal structure on all scales, and are typically
clumpy or filamentary. Direct imaging of CO in nearby
clouds reveals structure on all scales down to lengths of $\sim 0.01$ pc
and masses of $\sim 0.01 \, {\rm M}_{\sun}$ (Falgarone, Puget,
\& P\'erault 1992; Langer et al. 1995). Studies of the time variability
of absorption lines indicates the presence of structure in the
dense gas on scales down to lengths of $\sim 5 \times 10^{-5}$ pc and
masses of $\sim 5 \times 10^{-9} \, {\rm M}_{\sun}$ (Marscher, Moore
\& Bania 1993; Moore \& Marscher 1995).

In this hierarchy of sizes, however, not all scales are of equal
interest to astronomers. Stars form by gravitational collapse of
dense regions within molecular clouds; much interest is therefore
focused on the scale corresponding to the mass of protostars.
Surveys of nearby clouds (within 500 parsecs of the earth)
have focused attention on dense cores with typical diameters of
$\sim 0.1 \, {\rm pc}$, masses of $\sim 30 \, {\rm M}_{\sun}$,
and densities of $\sim 2 \times 10^4 \, {\rm cm}^{-3}$
(Myers, Linke, \& Benson 1983; Benson \& Myers 1989). In addition
to dense cores embedded within GMCs, our galaxy also contains
isolated dense clouds known as Bok globules (Bok \& Reilly 1947).
Neither dense cores nor Bok globules are spherical, as a general
rule. The projected axis ratio,
$q \equiv b/a$, of small Bok globules has an average value
$< \! q \! > \approx 0.6$ (Clemens \& Barvainis 1988, hereafter CB; Bourke,
Hyland, \& Robinson 1995; hereafter BHR). Embedded dense cores are similarly
flattened, with $ < \! q \! > \approx 0.5 - 0.6$ (Myers et al. 1991;
Tatematsu et al. 1993).

The projected axis ratios of globules and cores are of interest
because they place constraints on the intrinsic shapes of
these objects. The intrinsic shapes of isolated globules and embedded
cores are determined by a variety of physical processes. Globules
and cores are sculpted by the self-gravity of the gas which they
contain and by thermal pressure, turbulent pressure, and magnetic
pressure. Since star formation occurs in these dense regions,
their physical properties are further modified by outflows
and winds from the protostars which may be embedded within them.
Recent optical surveys of Bok globules (CB; BHR;
Hartley et al. 1986) and millimeter surveys of dense cores (Benson
\& Myers 1989; Loren 1989; Lada, Bally, \& Stark 1991; Nozawa et al.
1991) provide data sets of apparent axis ratios. As a cautionary note,
however, it should be pointed out that a core or globule is not
a solid, well-defined object with easily measurable axis ratios. It
is merely one scale in a hierarchy of structure. The shape of a dense core
as defined by the emission of one molecular line, moreover, will not be
precisely the same as its shape defined by the emission of a
different molecule. Readers are advised to regard the shapes measured
in surveys with a certain amount of skepticism.

Previous studies (David \& Verschueren 1987; Myers et al. 1991; Fleck 1992) 
have placed constraints on the permitted intrinsic shapes of dense cores.
For instance, the mean apparent axis ratio $< \! q \! > \approx
0.5 - 0.6$ for cores is fully consistent
with a population of prolate objects, but only marginally
consistent with an oblate population (Myers et al. 1991;
Fleck 1992). More sophisticated analytic tools for examining
sets of axis ratios have now been developed by investigators
studying the intrinsic axis ratios of elliptical galaxies and
other stellar systems. For instance, starting with a nonparametric
estimate for the distribution $f(q)$ of apparent axis ratios,
it is possible to find an estimate for the distribution $N(\gamma)$
of intrinsic axis ratios, given the assumption that
the systems considered are either oblate or prolate
spheroids and are randomly oriented with respect
to the observer (Tremblay \& Merritt 1995; Ryden 1996).

Using this technique, one is able, for instance, to
test the hypothesis that all dense cores or Bok globules are
randomly oriented prolate objects.
A test of the randomly oriented prolate hypothesis, more stringent than simply looking
at the mean $< \! q \! >$, is
provided by examining the entire distribution $f(q)$.
I first compute an estimate $\hat f (q)$ for the distribution
of apparent axis ratios, then mathematically invert it to
find a estimated distribution ${\hat N}_P (\gamma)$ of intrinsic
axis ratios. The inversion is mathematically unique, but the
resulting function ${\hat N}_P$ is only physically meaningful
if it is non-negative for all values of $\gamma$.
If ${\hat N}_P$ dips below zero at a statistically
significant level, the initial hypothesis -- that the cores or
globules are randomly oriented prolate objects -- can be rejected.

Similarly, one can test the hypothesis that cores or globules
are randomly oriented oblate objects. The available kinematic information,
however, indicates that dense cores are not rotationally supported
oblate spheroids (Goodman et al. 1993). Large scale maps also show that
many dense cores are aligned with larger filamentary structures,
indicating that a prolate geometry is more likely in such cases
(Myers et al. 1991). Thus, it should not be surprising when the
randomly oriented oblate hypothesis turns out to be
emphatically rejected for samples of dense cores.

In section 2, I give a brief outline of the mathematical techniques
used to find the distribution of intrinsic axis ratios, given either
the prolate or oblate hypothesis -- a fuller description, for those
who desire it, is given
in Ryden (1996). In section 3, I apply these techniques
to Bok globules, and in section 4, I apply them to dense cores embedded
within molecular clouds. Generally speaking, for each data set considered,
the prolate hypothesis gives a better fit than does the oblate hypothesis,
given the assumption that the globules and cores are randomly oriented.
In section 5, I discuss the implications of this finding for the
origin and evolution of dense cores and globules.

\section{Methods}

I start with a sample $q_1$, $q_2$, $\dots$, $q_N$ measured axis ratios
for a sample of $N$ globules or dense cores.
The nonparametric kernel estimator for the underlying
distribution $f(q)$ is
\begin{equation}
{\hat f} (q) = {1 \over N h} \sum_{i=1}^N K \left( {q - q_i \over h}
\right) \ ,
\end{equation}
where $K$ is the kernel function. A discussion of kernel
nonparametric estimators, as applied to astronomical
problems, is given by Merritt \& Tremblay (1994), Vio et al. (1994),
and Merritt \& Tremblay (1995). For my purposes, I want
a smooth differentiable kernel, so I use a Gaussian:
\begin{equation}
K(x) = {1 \over \sqrt{2 \pi} } e^{-x^2/2}  \ .
\end{equation}
For a kernel width $h$, I adopt
\begin{equation}
h = 0.9 A N^{-0.2} \ ,
\end{equation}
where $A$ is the smaller of the standard deviation of the sample and
its interquartile range divided by $1.34$ (Silverman 1986; Vio et al. 1994).
For samples which are not strongly skewed, this formula for $h$ is
a good approximation to the value which minimizes the integrated
mean square error in $\hat f$. For the samples examined in this paper,
$0.06 < h < 0.10$.

Since $f(q)$ must be equal to zero for $q < 0$ and $q > 1$, reflective
boundary conditions are applied at the boundaries $q = 0$ and $q=1$
(Ryden 1996). The  use of reflective boundary conditions ensures
the proper normalization
\begin{equation}
\int_0^1 {\hat f} (q) dq = 1 \ .
\end{equation}
One drawback of reflective boundary conditions is that they compel
the slope of $\hat f$ to be equal to zero at the boundaries, which
might not be true of the actual distribution $f$. The shape of $\hat f$
within a distance $h$ of the boundaries should therefore be viewed
with skepticism. 

If the globules or cores in the sample are all randomly oriented
oblate spheroids, then the estimated distribution ${\hat N}_O ( \gamma )$
for the intrinsic axis ratio $\gamma$ is given by the relation
\begin{equation}
{\hat N}_O ( \gamma ) = {2 \gamma \sqrt{1 - \gamma^2 } \over \pi }
\int_0^\gamma {d \ \over dq} \left( {\hat f} / q \right) {dq \over
\sqrt{\gamma^2 - q^2} } \ .
\end{equation}
If the globules or cores are assumed to be randomly oriented prolate spheroids,
then the estimated distribution ${\hat N}_P ( \gamma )$ for the
intrinsic axis ratio is
\begin{equation}
{\hat N}_P ( \gamma ) = {2 \sqrt{ 1 - \gamma^2 } \over \pi \gamma }
\int_0^\gamma {d \ \over dq} \left( q^2 {\hat f} \right) {dq \over
\sqrt{\gamma^2 - q^2} } \ .
\end{equation}
If ${\hat N}_O$ is negative for any value of $\gamma$, at a statistically significant level,
then I can reject the hypothesis that the sample is
drawn from a population of randomly oriented oblate objects. Either
they are not all oblate, or they are not randomly oriented. Having
a negative number of objects with a given value of $\gamma$ is
permissible from a purely mathematical viewpoint, but is
nonsensical from a physical viewpoint. Similarly,
if ${\hat N}_P$ drops below zero at a statistically significant level,
the randomly oriented prolate hypothesis can be rejected. The statistical significance
of excursions below zero is estimated by bootstrap resampling techniques
(Scott 1992; Merritt \& Tremblay 1994). From the original data set
$q_1$, $q_2$, $\dots$, $q_N$, I draw, with replacement, a new set
of $N$ data points. Using this bootstrap resampling of the original
data, I compute new estimates for $\hat f$, ${\hat N}_O$, and
${\hat N}_P$. After making a large number $n$ of these bootstrap estimates
(for this paper, I used $n=800$), confidence intervals can be placed
on the original estimates $\hat f$, ${\hat N}_O$, and ${\hat N}_P$.
At each value of $\gamma$, for instance, an upper limit can be placed
on ${\hat N}_O$ at the 99\% one-sided confidence level,
by finding the value of ${\hat N}_O$ such that
1\% of the bootstrap estimates lie above this value. If this
upper confidence limit drops below zero, then the hypothesis that
the objects are randomly oriented oblate spheroids can be rejected
at the 99\% confidence level. Similar analysis of ${\hat N}_P$ can
be used to test the hypothesis that the objects in the original
sample are randomly oriented prolate spheroids.

The confidence intervals derived from bootstrap resampling represent
only the error resulting from finite sample size. Additional errors
in the estimates $\hat f$, $\hat N_O$, and $\hat N_P$ will be
present as a result of errors in the measured values of $q$. If
a typical measurement error $\sigma_q$ is smaller than the smoothing length
$h$, then the effects of measurement error can be ignored. The
errors in the axis ratios of dense cores observed at millimeter wavelengths
are relatively small, if the effects of the beamwidth are correctly
subtracted (see section 4 below). However, the errors in the
published axis ratios for surveys of Bok globules (CB; BHR)
are significantly larger than $h$.
A further complication is added by the fact
that errors in $q$ are generally not Gaussian, so they cannot
simply be added in quadrature to the kernel width $h$. The presence of
non-Gaussian errors in the measurement of the axis ratio of nearly
circular objects can significantly affect the shape of $\hat f$ when
$q \sim 1$. This distortion of $\hat f$, in turn, can affect the
conclusions which are drawn about the intrinsic shapes of the
observed objects. A cautionary tale is related by Franx \& de Zeeuw
(1992), who deduced the intrinsic ellipticity $\epsilon$ of disk
galaxies from their observed axis ratios. Assuming Gaussian errors
in $q$, their best model had $\epsilon = 0.06$; however, the introduction
of non-Gaussian errors led to a best fitting model with $\epsilon = 0$.
In the following section, it will also be seen that the introduction of
non-Gaussian rounding errors distorts the distribution $\hat f (q)$ for
Bok globules.

\section{Isolated Bok Globules}

Bok globules are not, in general, spherical. Bok \& Reilly (1947) defined
globules as ``approximately circular or oval dark objects of small size'',
as contrasted to the ``wind-blown wisps of dark nebulosity'' which can
also be found in the interstellar medium of our galaxy. CB
constructed a catalog of small Bok globules
with declination $\delta > -36\deg$. They searched Palomar Observatory
Sky Survey (POSS) plates for isolated, opaque globules with diameters less
than $10'$. The axis ratio of the globules was explicitly not a
selection criterion. Their complete catalog contains 248 objects,
with a mean angular size of $4'$. The apparent axis ratio of each
globule was determined by approximating its shape as an ellipse,
and then measuring the minor and major axes of the fitted ellipse.
The axis ratios measured for the complete sample ranged from
$q = 1.00$ to $q = 0.14$, with a mean $< \! q \! > = 0.59$ and
standard deviation $\sigma_q = 0.23$.

To begin the analysis, I na\"{\i}vely take the values for the
axis length $a$ and $b$ published by CB,
and use the resulting values of $q = b/a$ to find the estimated
distribution ${\hat f} (q)$. The function $\hat f$ determined in
this way is shown as the solid line in the upper panel of Figure
1. In this Figure (and in every subsequent Figure in this paper),
the dashed lines indicate the 80\% confidence band and
the dotted lines show the 98\% confidence band, as determined
by bootstrap resampling. The estimated distribution ${\hat N}_O$ of
intrinsic axis ratios, given the randomly oriented oblate hypothesis, is shown in
the middle panel of Figure 1. The estimated distribution ${\hat N}_P$
of axis ratios, given the randomly oriented prolate hypothesis, is shown in the bottom panel. The
most striking aspect of the estimate $\hat f (q)$ found in this
manner is that it is bimodal. In addition to a broad maximum around
$q \sim 0.55$, there is a second peak at $q = 1$. Does this bimodality
indicate that there are two populations of globules, one flattened
and one nearly spherical? No. The peak at $q = 1$ is an artifact,
the result of rounding errors in $q$. CB, in measuring
the axis lengths of globules on the POSS plates, rounded $a$ and $b$
to the nearest millimeter, corresponding to $1{\farcm}12$ in
angular scale. Many of the globules in the survey are quite small;
42 out of 248 have $a \leq 2 \, {\rm mm}$. Consequently, rounding
of the axis lengths can have a large effect on the measured value
of $q$. For instance, a globule whose true size on the POSS plate
is $1.6 \, {\rm mm} \times 2.4 \, {\rm mm}$ will be tabulated as
being $2.0 \, {\rm mm} \times 2.0 \, {\rm mm}$. Its axis ratio will
be erroneously computed as $q = 1.00$ instead of its true value of
$q = 0.67$. A major effect of rounding errors is that small flattened
globules will be incorrectly classified as being circular.

The effect of the non-Gaussian rounding errors can be approximately
compensated for. To the values of $a$ and $b$ tabulated by
CB for each globule, I add an error term $\Delta$
drawn uniformly from the interval $ -0{\farcm}56 < \Delta < 0{\farcm}56$.
I then compute the new values of $q$ after the error terms are added on,
and compute ${\hat f} (q)$. After repeating this process 800 times, with
a different seed for the random number generator each time, I take
the average value of the 800 estimates $\hat f (q)$ as the new best
estimate of the underlying distribution $f (q)$, taking into
account the errors introduced by rounding the axis lengths $a$ and
$b$. The best estimate found in this way is given as the solid line
in the top panel of
Figure 2. The 80\% confidence bands (dashed lines) and the 98\% confidence bands
(dotted lines) include
both the errors due to finite sample size and the errors in
$q$ due to rounding.

Note that when the rounding errors are compensated for, the peak at
$q = 1$ disappears. The distribution $\hat f$ now has a single maximum
at $q = 0.6$. The mean value of $q$ is $< \! q \! > = 0.57$ and the
standard deviation is $\sigma_q = 0.22$. The derived values of ${\hat N}_O$,
given the oblate hypothesis, and ${\hat N}_P$, given the prolate hypothesis,
are shown in the middle panel and bottom panel of Figure 2. The oblate
hypothesis cannot quite be rejected at the 99\% one-sided confidence level,
but can be rejected at lower confidence levels. If the globules are
all oblate spheroids, then they must be quite flattened. Few or no
globules, if they are oblate, can have $\gamma \gtrsim 0.7$, and
the peak in ${\hat N}_O$ is at $\gamma \approx 0.25$. The observed
axis ratios, when corrected for rounding errors, are fully consistent
with the hypothesis that globules are all prolate objects with
random orientations. The
distribution ${\hat N}_P$ (bottom panel of Figure 2) is everywhere
positive, yielding a mean axis ratio $< \! \gamma \! > = 0.48$ and
standard deviation $\sigma_\gamma = 0.20$. The prolate hypothesis
yields a broader range of intrinsic axis ratios than does the
oblate hypothesis, with prolate globules ranging in shape from
filamentary structures with $\gamma \sim 0.1$ to nearly spherical
globules with $\gamma \sim 1$. 

The catalog of CB in the northern sky
is complemented by that of BHR in the southern
sky. Starting with the comprehensive catalog of Hartley et al. (1986),
which lists dark clouds with $\Delta < - 33\deg$, BHR
selected highly opaque clouds with $a < 10'$. A total of 169 globules
met their selection criteria. The axis lengths of the selected
globules were measured from SERC Schmidt survey plates. Axis lengths
longer than $3'$ were rounded to the nearest arcminute; axis lengths
shorter than $3'$ were rounded to the nearest $0.5$ arcminute. This
rounding of axis lengths was sufficient to create significant errors
in the axis ratio $q$. Consequently, I computed the estimated distribution
function ${\hat f} (q)$ using the correction technique described
above. To axis lengths less than $3'$, I added an error $\Delta$
drawn uniformly from the interval $-0{\farcm}25 < \Delta < 0{\farcm}25$.
For axis lengths greater than $3'$, $-0{\farcm}5 < \Delta < 0{\farcm}5$.
The value of ${\hat f}$, including corrections for rounding errors,
is given in the upper panel of Figure 3. The 80\% and 98\% confidence bands
include the effects both of finite sample size and of rounding errors.

The BHR globules are slightly rounder on average than
the CB globules; $< \! q \! > = 0.61$ as compared
to $< \! q \! > = 0.57$. The standard deviation for each sample
is $\sigma_q = 0.22$. In addition, the $\hat f$ for the BHR
globules lacks the dip at $q \approx 1$ which is seen in
$\hat f$ for the CB sample. The difference in apparent shapes
for the CB sample and the BHR sample is a result of the
difference in their definitions of a Bok globule, and not
a reflection of differences in the properties of Bok globules
in the northern and southern hemispheres.

The oblate distribution ${\hat N}_O$ for the BHR sample is shown
in the middle panel of Figure 3; it is consistent with the hypothesis that
the globules are randomly oriented oblate spheroids. If so, however,
they must be intrinsically quite flat, with a mean and standard deviation
of $< \! \gamma \! > = 0.33$ and $\sigma_\gamma = 0.15$; few or no
globules can have $\gamma \gtrsim 0.7$. If, by contrast, the globules
are randomly oriented prolate spheroids, they will have a broader
range of intrinsic axis ratios, shown in the bottom panel of Figure 4.
Under the prolate hypothesis, the mean and standard deviation of
the intrinsic axis ratio are $< \! \gamma \! > = 0.52$ and $\sigma_\gamma
= 0.21$.

In conclusion, both the BHR and CB data sets show that the spherical
approximation is a poor one for Bok globules; they must be intrinsically
quite flat. The hypothesis that Bok globules are randomly oriented
oblate objects cannot be excluded. If the globules are oblate, however,
they must be quite flat, with a distribution peaking at $\gamma \sim 0.3$,
and equal to zero at $\gamma \gtrsim 0.7$. If the Bok globules are
randomly oriented prolate objects, they have a wider range of intrinsic
axis ratios. Still, there are few nearly spherical globules, even under
the prolate hypothesis. For the BHR sample, 21\% of prolate globules
have $\gamma \geq 0.7$; for the CB sample, only 14\% of the prolate
globules have $\gamma \geq 0.7$. 

\section{Embedded Dense Cores}

Bok globules, according to Bok's original definition,
are relatively well-defined and isolated objects. The
CB and BHR catalogs of Bok globules contain only small globules. Their
upper size limit of $10'$ corresponds to a linear scale of $0.7 \,
{\rm pc}$ at a distance of $500 \, {\rm pc}$. Small Bok globules
are of interest to astronomers because their relative simplicity makes
them good laboratories for testing theories of star formation.
Most star formation in our galaxy, however, occurs not within
simple Bok globules but within the more complex environment of
large molecular clouds. The dense cores of molecular clouds
(with ``dense'' meaning $n \gtrsim 10^4 \, {\rm cm}^{-3}$)
can be traced using the emission lines of CS (Lada et al. 1991)
or NH$_3$ (Benson \& Myers 1989). Each emission line probes
a different density and temperature. Cores defined by
CS emission have a larger velocity dispersion, density,
and mass than the cores defined by NH$_3$ emission
(Zhou et al. 1989). However, Myers et al.
(1991), by mapping dense cores in the lines of NH$_3$,
CS, and C$^{18}$O, have demonstrated that the maps in
the different lines are similar (but not identical)
in position, orientation, and elongation. Thus, it
seems that the observed shapes of cores are not strongly
dependent on the particular emission line which is chosen
for the purpose of observation.

Benson \& Myers (1989) made a search for dense cores in the
$(J,K) = (1,1)$ rotation inversion line of NH$_3$. They
surveyed 149 positions selected to be in the center of
opaque regions on POSS plates. The dark clouds selected
in this manner are mainly nearby, with 119 of the 149 clouds
known to be less than $500 \, {\rm pc}$ away. Of the 149 regions
surveyed, 66 were detected in NH$_3$. Of the regions where
NH$_3$ was detected, 37 were mapped, and had their apparent axis ratios
measured. Of the mapped regions, 26 were identified as being dense cores
embedded within a larger molecular complex (type C),
5 were identified as being isolated globules (type G),
and 6 were classified as type M, intermediate between types C and G.
A Kolmogorov-Smirnov (KS) test comparing the 26 type C cores to the
11 type M and G cores reveals no significant difference in their
distribution of apparent axis ratios ($P_{\rm KS}
= 0.86$). For the purposes of my analysis,
therefore, I will lump all the mapped regions together, and
refer to them as ``dense cores''. In addition to the dense cores
selected from POSS plates,
Benson \& Myers provided maps and axis ratios for an additional
4 dense cores, selected for observation because they were in the
same position as an IRAS point source. Estimated masses of the
41 mapped dense cores lie in the range $0.5 \, {\rm M}_{\sun} < M <
760 \, {\rm M}_{\sun}$, with a median of $13 \, {\rm M}_{\sun}$.

The upper panel of Figure 4 shows the
estimated distribution ${\hat f} (q)$ for the distribution
of apparent axis ratios for all 41 dense cores mapped by Benson
\& Myers (1989). The mean apparent flattening ($< \! q \! > = 0.59$) for
this sample is comparable to the mean for the BHR and CB samples
of Bok globules. However, the Benson \& Myers dense cores are
notably lacking in nearly circular objects; the maximum axis
ratio in the entire sample is $q = 0.89$. The lack of nearly
circular objects is the classic indication that a population
cannot consist of randomly oriented oblate spheroids. Indeed,
the oblate hypothesis can be rejected at a very high confidence level --
the middle panel of Figure 4 shows that ${\hat N}_O$
dips far below zero for $\gamma \gtrsim 0.8$. So great is the
paucity of nearly circular cores that even the randomly oriented prolate hypothesis
can be rejected at the 99\% confidence level (see the bottom
panel of Figure 4). In short, the Benson \& Myers sample of dense
cores cannot be a population of randomly oriented axisymmetric
objects. Either they contain a significant number of triaxial cores
or there is a selection effect which excludes axisymmetric cores from the
sample when their axis of symmetry is close to the line of sight.
Compared to populations of axisymmetric objects, populations of
triaxial objects provide a better fit to samples with a lack
of apparently circular objects. This is because triaxial objects
appear to be nearly circular from only a very limited range of
viewing angles (Binggeli 1980; Binney \& de Vaucouleurs 1981).

Of the dense cores mapped by Benson \& Myers (1989), 21 have IRAS
point sources within one core diameter of the core's center. The
IRAS point sources are most plausibly explained as being
protostars or young stars embedded within the dense core.
It is not absurd to speculate that dense cores with embedded
IRAS sources could have different shapes, on average, than
cores without IRAS sources. For instance, cores of a particular
shape might be more efficient at collapsing to form stars. In
addition, an embedded protostar, in pouring out matter and radiation,
might change the shape of the core which encloses it. However,
it turns out that there is no significant difference between the
distribution of $q$ for the 21 cores with embedded IRAS sources and
the distribution for the 20 cores without IRAS sources. A KS
test comparing the two distributions yields $P_{\rm KS} = 0.9987$,
indicating a remarkably close similarity between the two distributions.
The similarity in axis ratios occurs in spite of the fact that
the cores with embedded sources have significantly larger radii,
masses, and line widths than the cores without sources (Benson
\& Myers 1989). Perhaps there is a preferred range of shapes for
NH$_3$ cores, which does not depend on the size of the core or
whether it contains an embedded protostar.

The dense cores mapped by Benson \& Myers exist lie within many
separate molecular clouds, with a wide range in galactic longitude.
To eliminate the variance which may exist between the cores of
different molecular clouds, with different physical and chemical
properties, it is useful to look at the population of dense cores
embedded within a single molecular cloud. Lada et al. (1991) studied
the dense cores of Lynds 1630 (alias Orion B), a molecular cloud
containing $\sim 10^5 \, {\rm M}_{\sun}$ of
gas (Maddalena et al. 1986), located $\sim 400 \, {\rm pc}$
away (Hilton \& Lahulla 1995).
Lada et al. mapped the cloud using the $J = 2 \to 1$ transition
of CS, using a beamwidth of $1{\farcm}8$ (corresponding to $0.21 \, {\rm pc}$).

Lada et al. (1991) identified individual ``clumps'' within Lynds 1630.
A clump was identified as a closed volume in position-velocity space,
containing CS emission at a $5\sigma$ level above the noise. The
entire region surveyed contained 42 clumps, as defined in this manner.
Shapes of individual clumps were determined by measuring the major
and minor axis lengths of the $5\sigma$ contours on the sky, subtracting
the beamwidth in quadrature. The majority of the clumps were unresolved,
but measured axis ratios were provided for 19 clumps with corrected major axis
length greater than the beamwidth ($2 a > 1{\farcm}8$). Virial masses
for these clumps lie in the range $8 \, {\rm M}_{\sun}  < M < 460 
\, {\rm M}_{\sun}$, with a median of $104 \, {\rm M}_{\sun}$. Although 19
is not a large number, and consequently the derived error bands on
$\hat f$ and $\hat N$ are large, it is still interesting to compare
the shapes of the Lynds 1630 clumps to the more eclectic dense cores of Benson
\& Myers (1989). The 19 cores of Lada et al. (with a mean and standard
deviation in $q$ of $0.55 \pm 0.19$) have a distribution of apparent
shapes similar to that of the 41 dense cores of Benson \& Myers (with
a mean and standard deviation of $0.59 \pm 0.19$). A KS test comparing
the two distributions yields $P_{\rm KS} = 0.70$.
The nonparametric estimate ${\hat f} (q)$ for the Lada et al. clumps
is shown in the upper panel of Figure 5.
Because of the small number
of clumps, the oblate hypothesis cannot be excluded at the 99\%
one-sided confidence level (see the middle panel of Figure 5).
However, if the clumps are oblate and randomly oriented, then
they must all be quite flat, just as in the case of the Bok
globules studied by BHR and CB. The best-fitting function
${\hat N}_O$ peaks at $\gamma \sim 0.4$, and there can be
few or no oblate clumps with $\gamma \gtrsim 0.6$. Conversely,
the randomly oriented prolate hypothesis, as illustrated in the bottom panel of Figure 5,
is statistically more acceptable, and yields a broader range of
intrinsic axis ratios. The best-fitting function ${\hat N}_P$ yields
a mean axis ratio $< \! \gamma \! > = 0.46$ and a standard deviation
$\sigma_\gamma = 0.17$.

Another molecular cloud which has been surveyed for dense cores is
the $\rho$ Ophiuchus cloud. Its center is located at approximately
$b = 17\deg$, $l = 355\deg$ and is at a distance of $160 \, {\rm pc}$ from
the sun. Its total mass is estimated to be $\sim 3000 \, {\rm M}_{\sun}$
(Loren 1989). Thus, although smaller than Orion B, the $\rho$ Oph
cloud is closer, and can be mapped at higher resolution. Loren (1989)
mapped the $\rho$ Oph cloud using the $J = 1 \to 0$ transition of
$^{13}$CO. The telescope beamwidth was $2{\farcm}4$ FWHM, corresponding
to a spatial resolution of $0.11 \, {\rm pc}$.
The catalog of Loren (1989) contains 89 ``clumps'', located at
maxima of the emission in velocity-position space. Each clump
must be clearly separated from other clumps in order to be
included in the catalog. The masses estimated from
the integrated $^{13}$CO intensity lie in the range
$0.6 \, {\rm M}_{\sun} < M < 850 \, {\rm M}_{\sun}$, with a median
of $8.8 \, {\rm M}_{\sun}$. The major and minor axis lengths of
each clump are the FWHM along the principal axes, projected
onto the sky (the beamwidth is subtracted in quadrature).
The definition of a ``clump'' and of its major and minor axis lengths
thus differ from the definitions used by Lada et al. (1991); be
cautious in comparing the two different types of clump.

The estimated distribution ${\hat f} (q)$ for the
Loren clumps (upper panel of Figure 6) is marked by a
scarcity of objects with $q \gtrsim 0.8$. Of 89 clumps,
only 6 have $q > 0.8$. The distribution of $q$ for the
clumps of Loren (1989) is significantly different 
at the 98\% confidence from that for 
the dense cores of Benson \& Myers (1989); a K-S test yields
$P_{\rm KS} = 0.015$. The Loren clumps are flatter, on
average, than the Benson \& Myers cores, with a mean
and standard deviation in $q$ of $0.48 \pm 0.20$.
The difference between the clump shapes
found by Loren (1989) and those found by Lada et al. (1991)
is not statistically significant: $P_{\rm KS} = 0.41$.
Because of the lack of nearly circular clumps in the
Loren sample, the randomly oriented oblate hypothesis can be rejected
at the 99\% confidence level; see the middle panel of Figure 6.
The prolate hypothesis, however, is statistically acceptable,
as shown in the bottom panel of Figure 6. If the clumps
are randomly oriented prolate objects, they must be quite flat,
with $< \! \gamma \! > = 0.39$ and $\sigma_\gamma = 0.17$, and with
very few clumps having $\gamma > 0.7$.

The dense cores and clumps observed by the various surveys differ
somewhat in their definitions. Their masses, however, all lie
in the range $0.5 \, {\rm M}_{\sun} < M < 900 \, {\rm M}_{\sun}$, and
their characteristic density is $\sim 10^4 \, {\rm cm}^{-3}$.
Broadly speaking, the dense
cores and clumps are more readily explained by the prolate
hypothesis than by the oblate hypothesis. This conclusion results from
the lack of nearly circular objects in the samples. The cores
and clumps must be intrinsically quite nonspherical; if they
are randomly oriented prolate objects, the mean axis ratio
is approximately $\gamma \sim 0.4$.

In focusing solely on the
dense cores of molecular clouds, however, I have been neglecting
the larger structures within clouds. Molecular clouds show
substructure on all resolvable scales, including scales larger
and smaller than individual dense cores. The existence
of hierarchical structure on such a
broad range of scales has prompted investigators
to describe molecular clouds in terms of their fractal
dimension (Dickman, Horvath, \& Margulis 1990; Falgarone,
Phillips, \& Walker 1991). However, different
physical processes dominate at different length scales, which makes
it unlikely that the structure is a perfectly self-similar fractal.
To investigate the dependence of the shape of substructure
upon size, I can use the molecular survey of Nozawa et al. (1991),
who identified and cataloged substructures of different size
in their survey.

Nozawa et al. (1991) mapped an area
in the Ophiuchus region using the $J = 1 \to 0$ transition
of $^{13}$CO. The telescope beamwidth was $2{\farcm}7$ FWHM.
The area mapped ($b = 15\deg \to 33\deg$; $l =
355\deg \to 12\deg$) is north of the $\rho$ Ophiuchus main cloud
mapped by Loren (1989). Nozawa et al. picked out two types of
substructure in their survey. The larger substructures, which
they call ``clouds'', are defined as the regions enclosed by
the $3\sigma$ contours in the integrated $^{13}$CO map. The
map contains 23 clouds defined in this manner, with estimated
masses in the range $13 \, {\rm M}_{\sun} < M < 1200 \, {\rm M}_{\sun}$.
The median cloud mass was $76 \, {\rm M}_{\sun}$. The smaller
substructures, called ``cores'', are defined as maxima in the
integrated $^{13}$CO emission. Thus, each cloud contains at least
one core; the largest cloud contains 8 separate cores.
The total number of cores mapped by Nozawa et al. (1991)
is 51 (including 3 cores not well resolved at $2{\farcm}7$).
The axis ratios of the cores were estimated by finding the
contour at the half peak intensity level and fitting the
contour with an ellipse.
The estimated core masses lie in the range $3 \, {\rm M}_{\sun}
< M < 160 \, {\rm M}_{\sun}$, with a median mass of $13
\, {\rm M}_{\sun}$. The mass of each core is from 0.6\% to
48\% of the mass of the cloud in which it is embedded.

The ``cores'' in the $^{13}$CO map of Nozawa et al. (1991) are
slightly larger, on average, than the ``clumps'' in the $^{13}$CO
map of Loren (1989). The two populations, however, do not have
different distributions of axis ratios at a high
statistical confidence level. Comparison of the axis ratios of
the two populations, using a KS test, yields $P_{\rm KS} = 0.18$.
The most noticeable difference between the two populations is
that the cores of Nozawa et al. (1991) are rounder; they have
a mean and standard deviation in $q$ of $0.57 \pm 0.21$, as
compared to $0.48 \pm 0.20$ for the Loren (1989) clumps.

The best-fitting estimate ${\hat f} (q)$ for the cores of Nozawa
et al. (1991) is shown in the upper panel of Figure 7. Once again,
we see a dip in $\hat f$ as $q \to 1$. However, the dip is not
as deep or as wide as in the case of the Loren (1989) clumps, shown
in Figure 6. As a result, the randomly oriented oblate hypothesis cannot be rejected
at the 90\% confidence level for the Nozawa cores
(see the middle panel of Figure 7). If the oblate hypothesis is
true, however, the cores must be very flat, with ${\hat N}_O$ peaking
at $\gamma \sim 0.3$, and with few or no cores having $\gamma \gtrsim
0.6$. The prolate hypothesis gives a better fit to the observations
(see the bottom panel of Figure 7). If the Nozawa cores are randomly
oriented prolate objects, then the mean axis ratio is $< \! \gamma
\! > = 0.48$, with a standard deviation of $\sigma_\gamma = 0.19$.
The results for the {\it cores} of Nozawa et al. (1991) are thus
similar to the results found for other dense cores and for Bok
globules; the prolate hypothesis gives a better fit than the
oblate hypothesis, and if the oblate hypothesis is insisted
upon, then the cores must all be quite flat. But what can be
deduced about the shapes of the {\it clouds}, which are larger
than the cores?

The clouds in the catalog of Nozawa et al. (1991) are distinctly
different in shape from the smaller, denser cores which are
embedded within them. A KS test comparing the 23 clouds with the
48 resolved cores yields $P_{\rm KS} = 2 \times 10^{-3}$. The
clouds are generally elongated filaments, with a mean and standard
deviation in their axis ratio of $< \! q \! > = 0.39$ and $\sigma_q
= 0.17$. The most elongated cloud, with $q = 0.07$, is also the 
most massive, with an estimated mass $M = 1200 \, {\rm M}_{\sun}$.
The estimated ${\hat f} (q)$ for the clouds, shown in the upper
panel of Figure 8, peaks at $q = 0.3$ and goes to zero at $q = 1$.
The severe lack of nearly spherical clouds means that the hypothesis
of randomly oriented oblate clouds can be strongly ruled out. The
99\% upper confidence band of ${\hat N}_O$
(seen in the middle panel of Figure 8) dips well below zero. The
roundest object in the sample of clouds has $\gamma = 0.7$. The
distribution of oblate objects which minimizes the number of apparently
circular objects is a population of infinitesimally thin disks ($\gamma = 0$);
this population, when placed at random orientations, produces a
distribution $f(q)$ which is uniform between $q = 0$ and $q=1$. The
probability that a sample of 23 infinitesimally thin disks, seen
at random angles, produces no projections with $q > 0.75$ is only
$P = (0.75)^{23} \approx 1.3 \times 10^{-3}$. Any other population
of oblate objects would produce an even smaller value of $P$.
The lack of nearly circular objects in the sample of clouds
can even be used to rule out the hypothesis that the clouds
are randomly oriented prolate objects. In the bottom panel of
Figure 8, showing ${\hat N}_P$, the one-sided 99\% confidence
level drops below zero for $\gamma > 0.84$. Thus, either the
clouds are intrinsically triaxial, or there is a selection effect
which eliminates prolate clouds whose axes of symmetry lie close
to the line of sight.

\section{Discussion}

Bok globules and dense molecular cloud cores, with densities
$n \gtrsim 10^4 \, {\rm cm}^{-3}$, are of particular interest
to astronomers because they are where stars form; see Myers (1985) and
Shu, Adams, \& Lizano (1987) for reviews. The assumption of
spherical symmetry in a star-forming dense core is seductive
from a theoretical viewpoint. It allows analytic calculations
of collapse, such as the self-similar isothermal solution of
Shu 1977), and permits numerical simulations with very high
resolution. However, the observations reveal that the spherical
assumption is a poor one. In the samples analyzed in this paper,
the mean apparent axis ratio ranges from $< \! q \! > = 0.48$
for the clumps observed by Loren (1989) to $< \! q \! > = 0.61$ for the Bok
globules cataloged by BHR. The intrinsic axis ratios are even
more elongated than the projected axis ratios. Under the prolate
hypothesis, the mean intrinsic axis ratio is $< \! \gamma \! > = 0.39$
for the Loren (1989) sample and $< \! \gamma \! > = 0.53$ for
the BHR sample.

For each of the samples analyzed, the prolate hypothesis (that all
cores and globules are randomly oriented prolate spheroids) yields
a better fit to the observed distribution of axis ratios than does
the oblate hypothesis. This conclusion agrees with other indications
that cores and globules are primarily prolate objects.
For instance, cores are unlikely to be rotating oblate
spheroids. In a sample of dense cores ($n
\gtrsim 10^4 \, {\rm cm}^{-3}$) studied by Goodman et al. (1993), only
29 of 43 had a statistically significant velocity gradient. The
measured gradients lie in the range $0.3 \to 4 \, {\rm km} \, {\rm
s}^{-1} \, {\rm pc}^{-1}$, with a typical ratio of rotational energy
to gravitational energy of only $\sim 0.02$. In addition, no
correlation was found between the rotation axis and the projected
short axis of the core, indicating that the cores are not rotationally
supported oblate objects. Moreover, a study of the few molecular clouds
with large velocity gradients (Arquilla \& Goldsmith 1986) shows
that clouds with significant gradients often have velocity fields
which are more complicated than would be produced by simple rotation.
If cores are shaped not by rotation but by anisotropic gravitational
fields (Fleck 1992), then the extremely elongated cores ($q \lesssim 0.3$)
must be prolate rather than oblate.
A study of the environment of dense cores offers hints that at least some
cores must be prolate. Of the 16 cores studied by Myers et al. (1991),
6 are observed to have their long axes aligned
with larger filamentary structures, which are almost certainly
prolate because of their extreme axis ratios ($q \sim 0.1$).

The tendency for dense cores to be embedded in long filaments
of lower density matter was noted by Schneider \& Elmegreen (1979),
who compiled a catalog of ``globular filaments'', defined as
long filamentary dark clouds with globule-like dense cores
spaced evenly along the filament. The distance between the
dense cores in a globular filament is equal to 3 times the
diameter of the cores. The work of Nozawa et al. (1991) confirms
the general picture of cores being prolate objects with
$q \sim 0.6$ embedded within larger, lower density,
more filamentary ($q \sim 0.4$) clouds. Although the long
axes of the dense cores tend to be aligned with the larger filaments
in which they are embedded (Myers et al. 1991), being embedded
within a larger structure does not greatly affect their shape.
Embedded cores have shapes similar to those of isolated globules.
Direct comparison of optically selected Bok globules
(BHR; CB) to dense cores detected at millimeter wavelengths
(Benson \& Myers 1989; Lada et al. 1991; Loren 1989; Nozawa et al. 1991)
is something that should only be done with caution. However,
the shapes of the embedded (type C) cores in the Benson \& Myers (1989)
sample are indistinguishable from the shapes of the isolated (type G) cores.

For all the samples of dense objects in this sample, whether they
are called globules, cores, or clumps, the best-fitting estimate
${\hat f} (q)$ dips below one as $q \to 1$. As a consequence,
${\hat N}_O ( \gamma )$ dips below zero, and the hypothesis that
the objects are randomly oriented oblate spheroids can be rejected
at some confidence level. For some samples -- the dense cores
of Benson \& Myers (1989) and the clouds of Nozawa et al. (1991) --
the best-fitting estimate ${\hat N}_P$ drops below zero.  For these
populations, we can reject the hypothesis of randomly oriented
prolate spheroids. Previously, I have concentrated on the implications
if the assumption of oblateness or prolateness is incorrect, and
the objects are actually triaxial. But what
if the other assumption I've made is incorrect, and the objects
in the sample are not randomly oriented? Neighboring cores on
a long filament, for instance, will tend to have their long
axes pointing in the same direction (Nozawa et al. 1991). This
correlation in the orientation angle will reduce the effective
number of independent viewing angles. Thus, the assumption of
independent random viewing angles is incorrect in this case,
and the error bands placed on $\hat f$, ${\hat N}_O$, and ${\hat N}_P$
by bootstrap resampling will be underestimates. Relaxing the
assumption that the orientations of cores are uncorrelated thus
makes it more difficult to rule out the hypotheses that the
cores are all oblate or all prolate.

It is not always obvious which assumption is incorrect when
a particular hypothesis is rejected. For instance, I rejected
the hypothesis that the clouds in the Nozawa et al. (1991) sample
are randomly oriented prolate spheroids. This rejection is based
on the lack of nearly circular clouds in the sample. Perhaps the assumption
of random orientations is incorrect, and prolate clouds whose
axis of symmetry lies close to the line of sight have been excluded
from the sample. An examination of the maps of the clouds (Nozawa et al.
1991) reveals, however, that the other assumption -- that the
clouds are intrinsically prolate, must be incorrect. In projection,
the clouds with the greatest apparent elongation are not ellipses,
but are arc-shaped. The major axis of the cloud, in some cases,
has a radius of curvature comparable to the axis length. This
curvature of filaments is also seen in the ``globular filaments''
of Schneider \& Elmegreen (1979). The long, filamentary clouds
must be intrinsically curved, which prevents them from appearing
exactly circular in projection from any angle.

On scales larger than dense cores, and with mean densities lower
than $n \sim 10^4 \, {\rm cm}^{-3}$, the distribution of molecular
gas is generally described as filamentary, wispy, or cobwebby, with
gas lying in long, curved filaments. Dense cores and Bok globules,
however, have less extreme axis ratios. It is not known which
forces are dominant in determining the shape of cores and globules.
It is possible, though, to state some forces which are not
greatly significant. Since the shape of cores and globules
is independent of whether they are embedded in larger structures
(Benson \& Myers 1989), external forces such as gravitational
tidal forces cannot determine the shapes in the majority of
cases. The shapes of cores also is independent of whether
they contain embedded protostars (Benson \& Myers 1989), so
outflows and radiation pressure from protostars cannot have
a significant effect on the overall shape of the core or
globule in which it is embedded. Since nearly all cores
are slowly rotating (Goodman et al. 1993), rotational flattening
does not play an important role. The dense cores are not
collapsing on a freefall timescale. The dynamical time for
a core with $n \sim 10^4 \, {\rm cm}^{-3}$ is $t_{\rm dyn} \sim
10^6 \, {\rm yr}$; the low efficiency of star formation in cores
and globules -- $\varepsilon \sim 0.3\%$ in the northern Ophiuchus
region, for instance (Nozawa et al. 1991) -- means that clouds
are contracting on timescales much longer than their dynamical
time. Thus, the ``pancaking'' effect of anisotropic collapse
(Lynden-Bell 1964; Lin, Mestel, \& Shu 1965)
is not what flattens cores and globules.

Dense cores and globules, it is generally proposed, are supported
by a combination of thermal pressure, turbulent pressure, and
magnetic pressure. For all but the smallest cores, the observed
line widths are greater than would be produced by thermal broadening
alone. Temperatures in dense cores and Bok globules are typically
$\sim 10 \, {\rm K}$. For $^{13}$CO, thermal velocities alone
produce a FWHM of $\Delta V = 0.13 \, {\rm km} \, {\rm s}^{-1}
( T / 10 \, {\rm K} )^{1/2}$. The observed width of $^{13}$CO
absorption lines is $\Delta v = 0.73 \to 2.24 \, {\rm km} \, {\rm s}^{-1}$
for the Lada et al. (1991) clumps and $\Delta v = 0.5 \to 1.8 \, {\rm km}
\, {\rm s}^{-1}$ for the Nozawa et al. (1991) cores. Supersonic
velocities are seen within Bok globules as well. When the BHR globules
were observed in NH$_3$ emission (Bourke et al. 1995),
the intrinsic linewidth in globules where
ammonia was strongly detected was $\Delta v = 0.28 \to 1.02 \, {\rm km} \,
{\rm s}^{-1}$, where thermal broadening would produce
$\Delta v = 0.16 \, {\rm km} \, {\rm s}^{-1}$ at a temperature
of $10 \, {\rm K}$. The additional broadening can be explained
as the result of turbulent motions within the globule or core
(Larson 1981). However, it seems that the shape of cores is
not crucially dependent upon the extent to which it is supported
by turbulence. Benson \& Myers (1989) found that the mean linewidth
for cores without embedded IRAS sources is $\Delta v = 0.27 \,
{\rm km} \, {\rm s}^{-1}$ is significantly narrower than the
mean width for cores with embedded sources, $\Delta v = 0.45 \,
{\rm km} \, {\rm s}^{-1}$, despite the fact that their kinetic
temperatures are all $\sim 10 \, {\rm K}$. Although the cores
with sources contain a higher fraction of turbulent
kinetic energy, their distribution of apparent shapes is
indistinguishable from that of cores without sources.

Magnetic forces can also play a significant role in determining
the structure of dense cores (McKee et al. 1993). Although measuring magnetic
fields in molecular clouds is a tricky business, available
measurements indicate that magnetic pressure and turbulent
pressure are of comparable importance in supporting cores
against collapse (Myers \& Goodman 1988). The presence of
magnetic flux is certainly capable of influencing the size and
shape of gaseous clumps. As an example, consider the scenario in
which dense cores form by gravitational condensation along a
filament -- a proposed method for the formation of globular
filaments (Schneider \& Elmegreen 1979; Elmegreen 1985). The
filament will break into globules whose separation is equal
to the wavelength $\lambda_m$ of the most rapidly growing
instability. In the absence of a magnetic field, $\lambda_m
= 5.4 D$, where $D$ is the initial diameter of the filament (Chandrasekhar
\& Fermi 1953).
If the filament is aligned with a magnetic flux, however,
$\lambda_m$ will be increased. In the real interstellar
medium, where the geometry of the magnetic field and the
gas distribution is far more complicated, the shape of
dense cores and Bok globules is determined by nonlinear
magnetohydrodynamic turbulence whose study will require
sophisticated numerical simulations. It is possible that the observed
shapes of Bok globules and dense cores will act as constraints
on future studies. It is also possible that the characteristic
distribution $f (q)$ which is observed, peaking at $q \sim 0.4 \to 0.6$
and falling below $f = 1$ as $q \to 1$, is a generic result
for a wide range of initial conditions and evolutionary histories.

\acknowledgments
I thank R. Barvainis, D. Clemens, R. Fleck, R. Pogge, and
K. Sellgren for their advice and assistance, and the onymous referee,
N. Evans, for his exceptionally helpful report.
This work was supported by NSF grant AST-9357396.

%
%

\begin{figure}
\caption{{\it Top}, nonparametric kernel estimate of the
distribution of apparent axis ratios for a sample of 248 northern
Bok globules (Clemens \& Barvainis 1988). The axis ratios
are {\it not} corrected for rounding errors. {\it Middle},
distribution of intrinsic axis ratios, assuming that
the globules are oblate. {\it Bottom}, distribution of
intrinsic axis ratios, assuming that the globules are
prolate. The solid line in each panel is the best estimate,
the dashed lines are the 80\% confidence band, found
by bootstrap resampling, and the dotted lines are the
98\% confidence band.
The kernel width is $h = 0.067$.
}

\caption{As Fig. 1, but the apparent axis ratios of
Clemens \& Barvainis (1988) are corrected
for rounding errors in the published catalog, using
the method described in the text.
The kernel width is $h = 0.063$.
}

\caption{As Fig. 1, but for a sample of 169 southern Bok
globules (Bourke, Hyland, \& Robinson 1995). The axis
ratios are corrected for rounding errors.
The kernel width is $h = 0.069$.
}

\caption{As Fig. 1, but for a sample of 41 dense cores
in nearby molecular clouds (Benson \& Myers 1989).
The kernel width is $h = 0.080$.
}

\caption{As Fig. 1, but for a sample of 19 dense cores, or ``clumps'',
in the Lynds 1630 (Orion B) molecular cloud (Lada, Bally,
\& Stark 1991).
The kernel width is $h = 0.093$.
}

\caption{As Fig. 1, but for a sample of 89 dense cores
in the $\rho$ Ophiuchus molecular cloud complex (Loren 1989).
The kernel width is $h = 0.074$.
}

\caption{As Fig. 1, but for a sample of 48 dense cores
in the Ophiuchus dark cloud complex (Nozawa et al. 1991).
The kernel width is $h = 0.090$.
}

\caption{As Fig. 1, but for a sample of 23 clouds in the
Ophiuchus dark cloud complex (Nozawa et al. 1991).
The kernel width is $h = 0.080$.
}

\end{figure}

\clearpage
\begin{figure}[h]
\epsfxsize=\hsize
\epsfbox{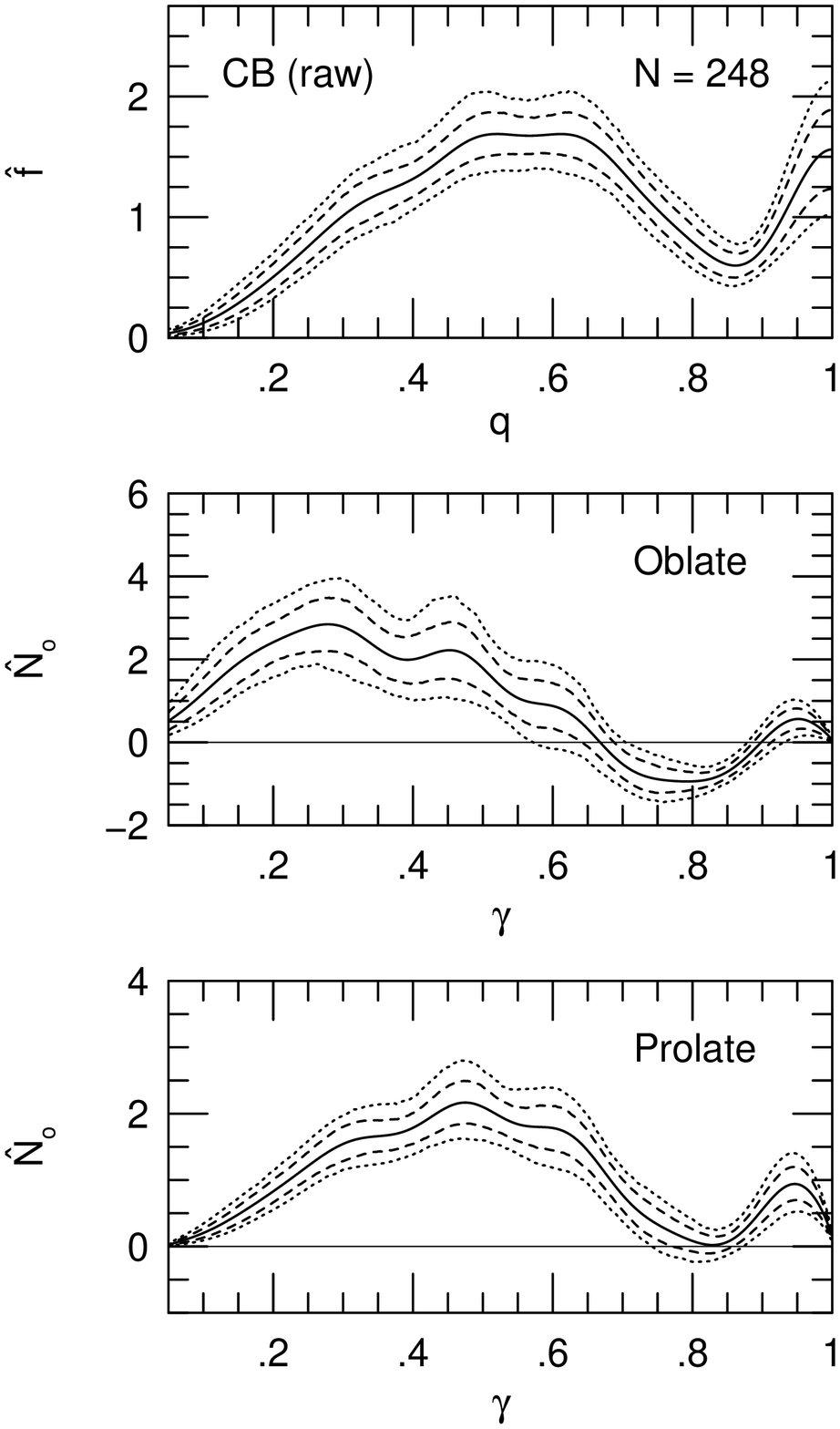}
\end{figure}

\begin{figure}[h]
\epsfxsize=\hsize
\epsfbox{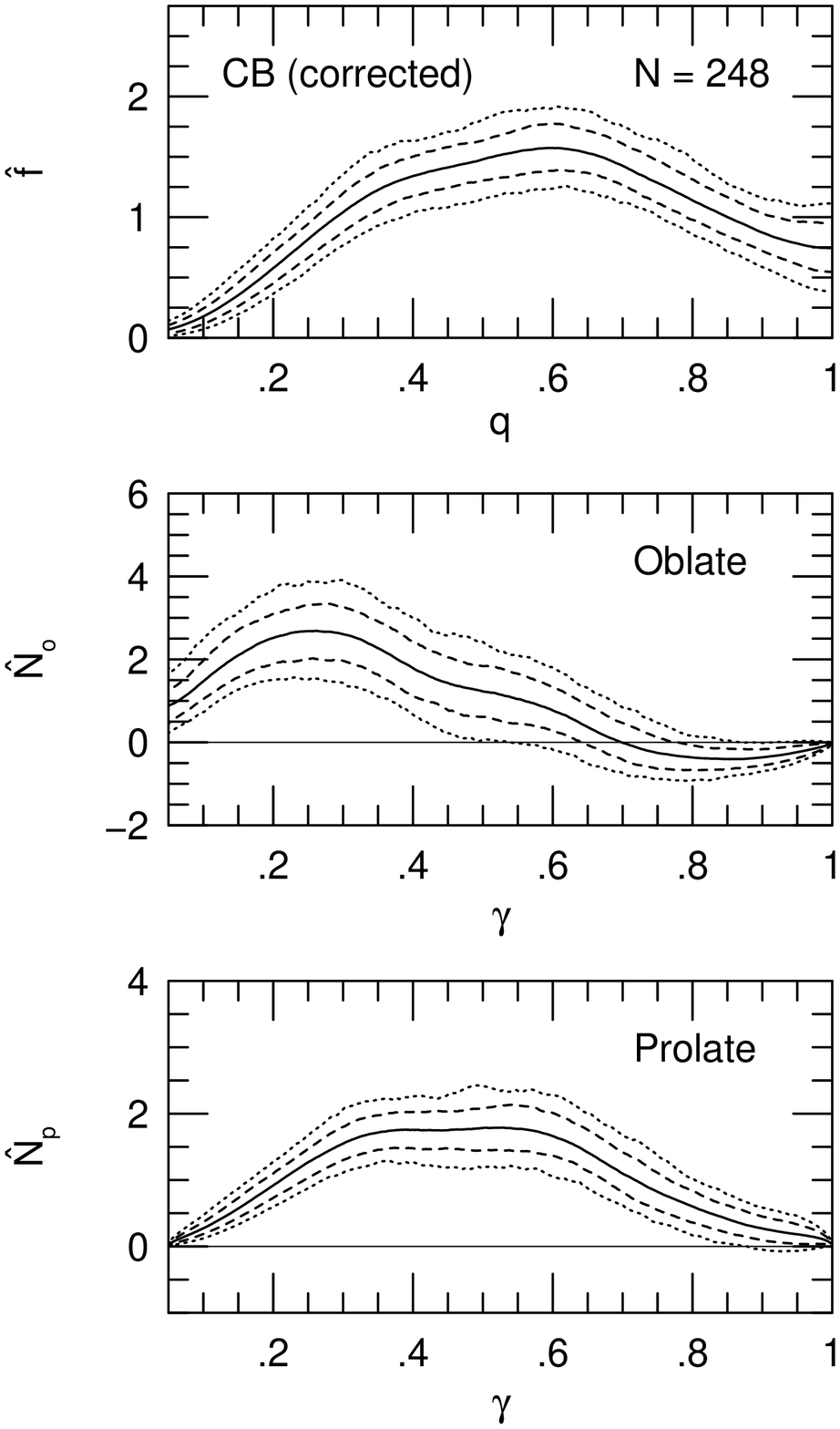}
\end{figure}

\begin{figure}[h]
\epsfxsize=\hsize
\epsfbox{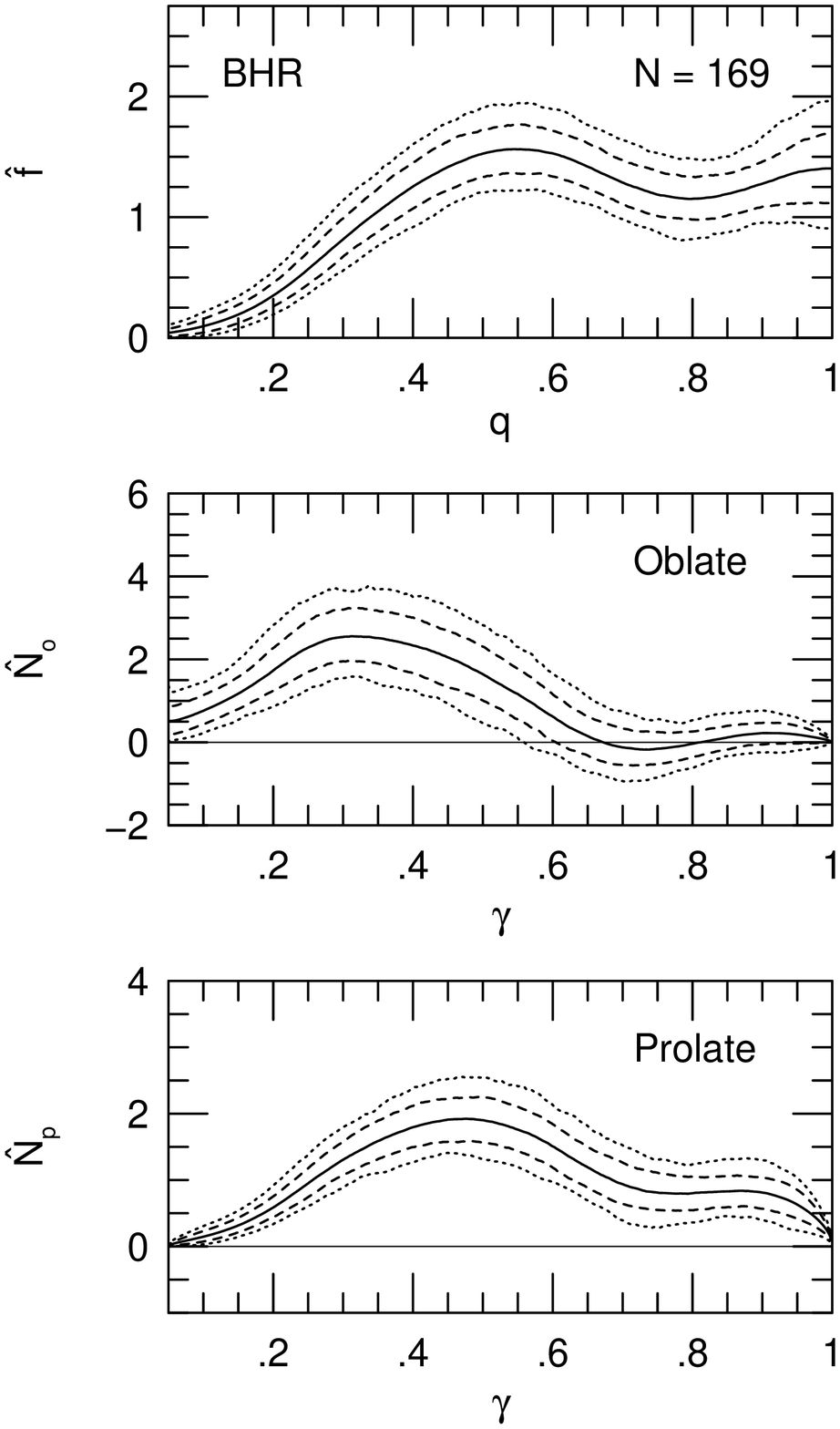}
\end{figure}

\begin{figure}[h]
\epsfxsize=\hsize
\epsfbox{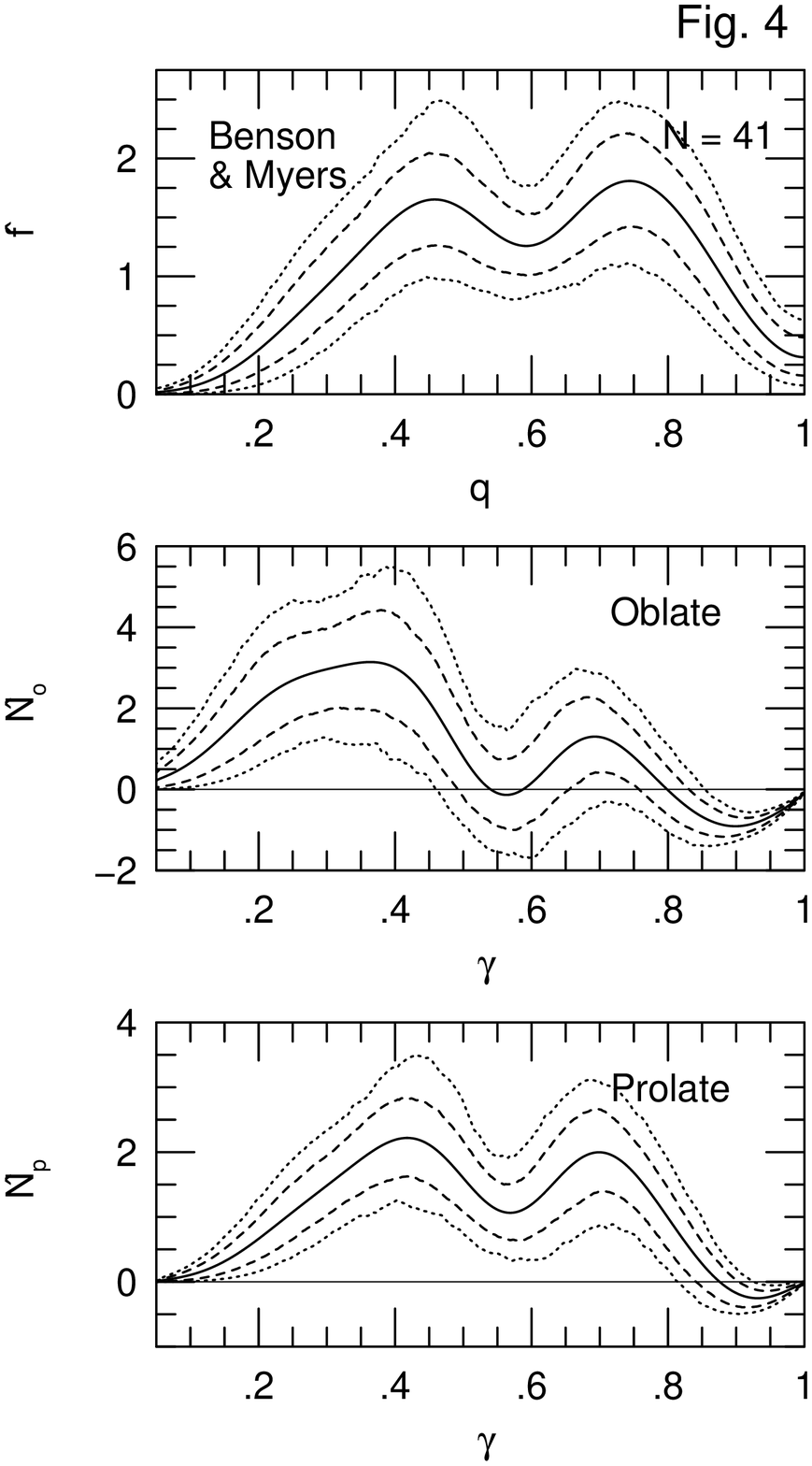}
\end{figure}

\begin{figure}[h]
\epsfxsize=\hsize
\epsfbox{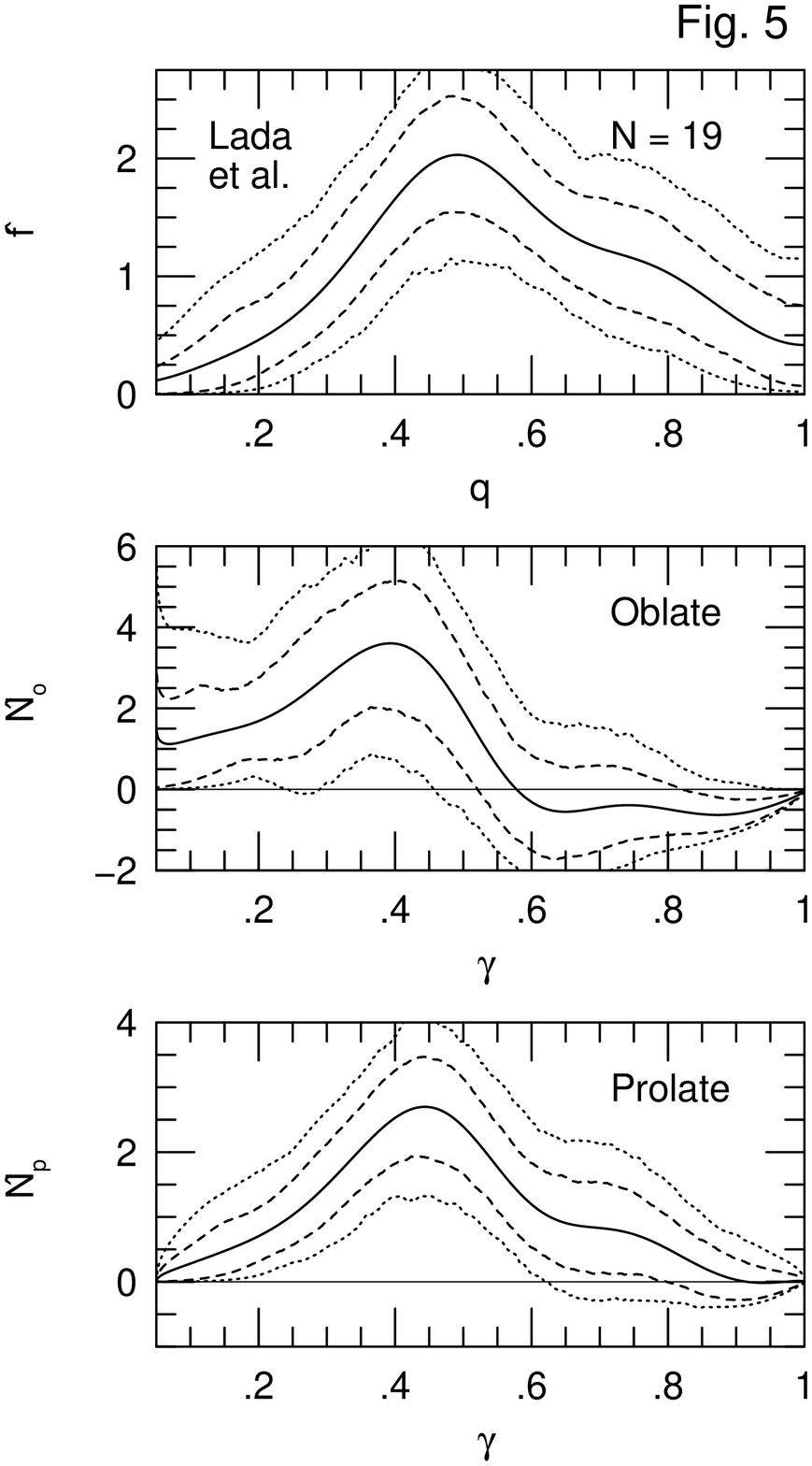}
\end{figure}

\begin{figure}[h]
\epsfxsize=\hsize
\epsfbox{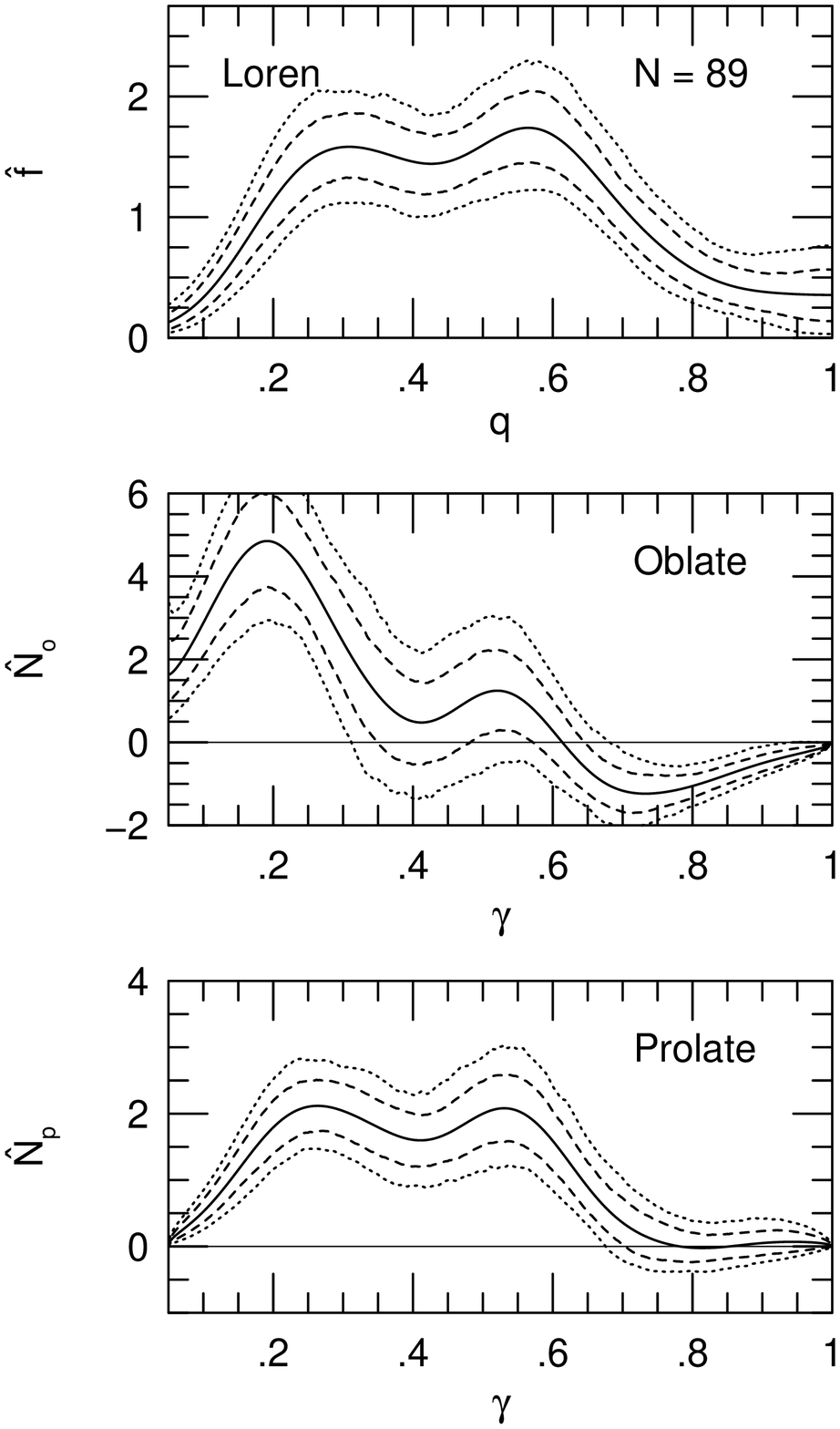}
\end{figure}

\begin{figure}[h]
\epsfxsize=\hsize
\epsfbox{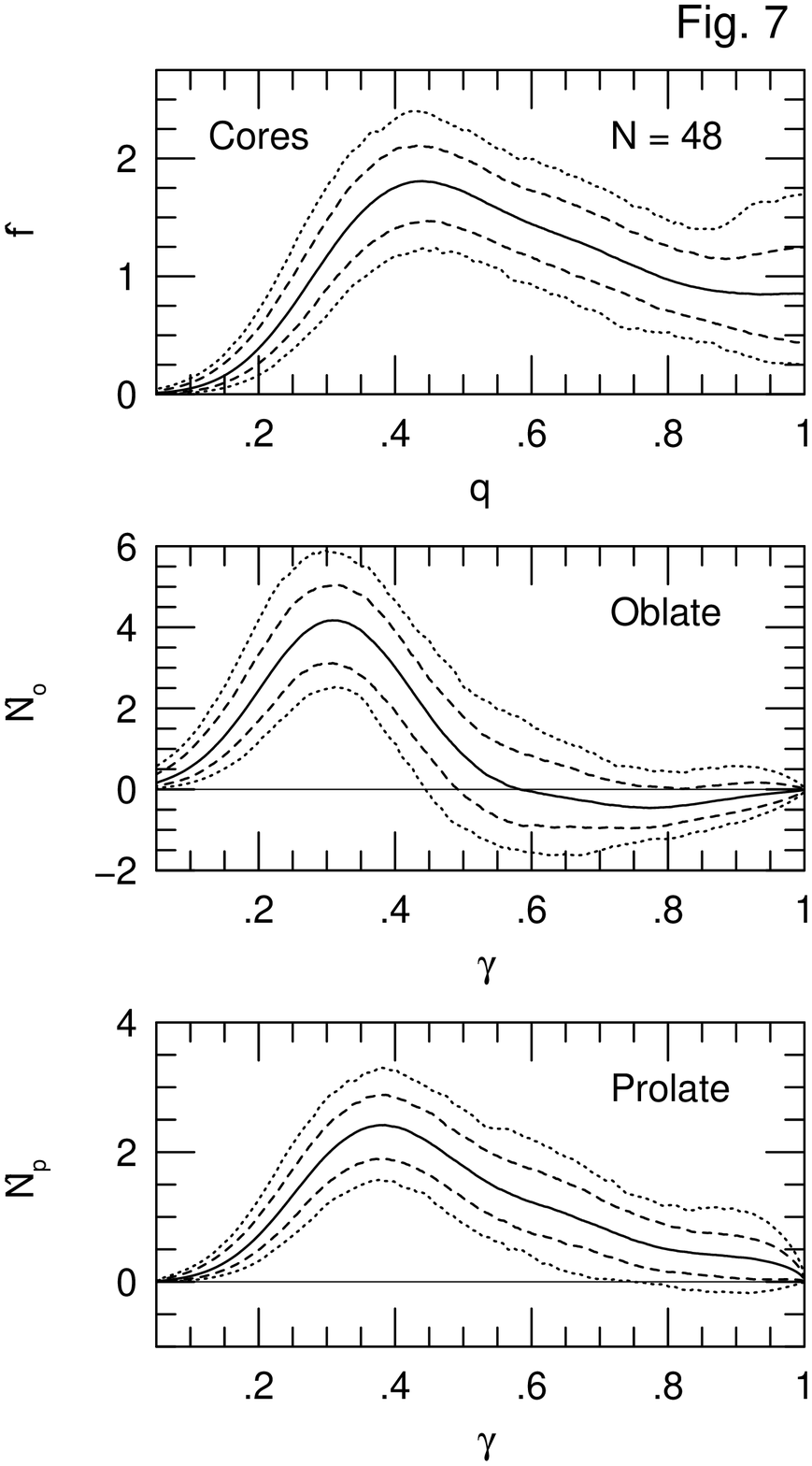}
\end{figure}

\begin{figure}[h]
\epsfxsize=\hsize
\epsfbox{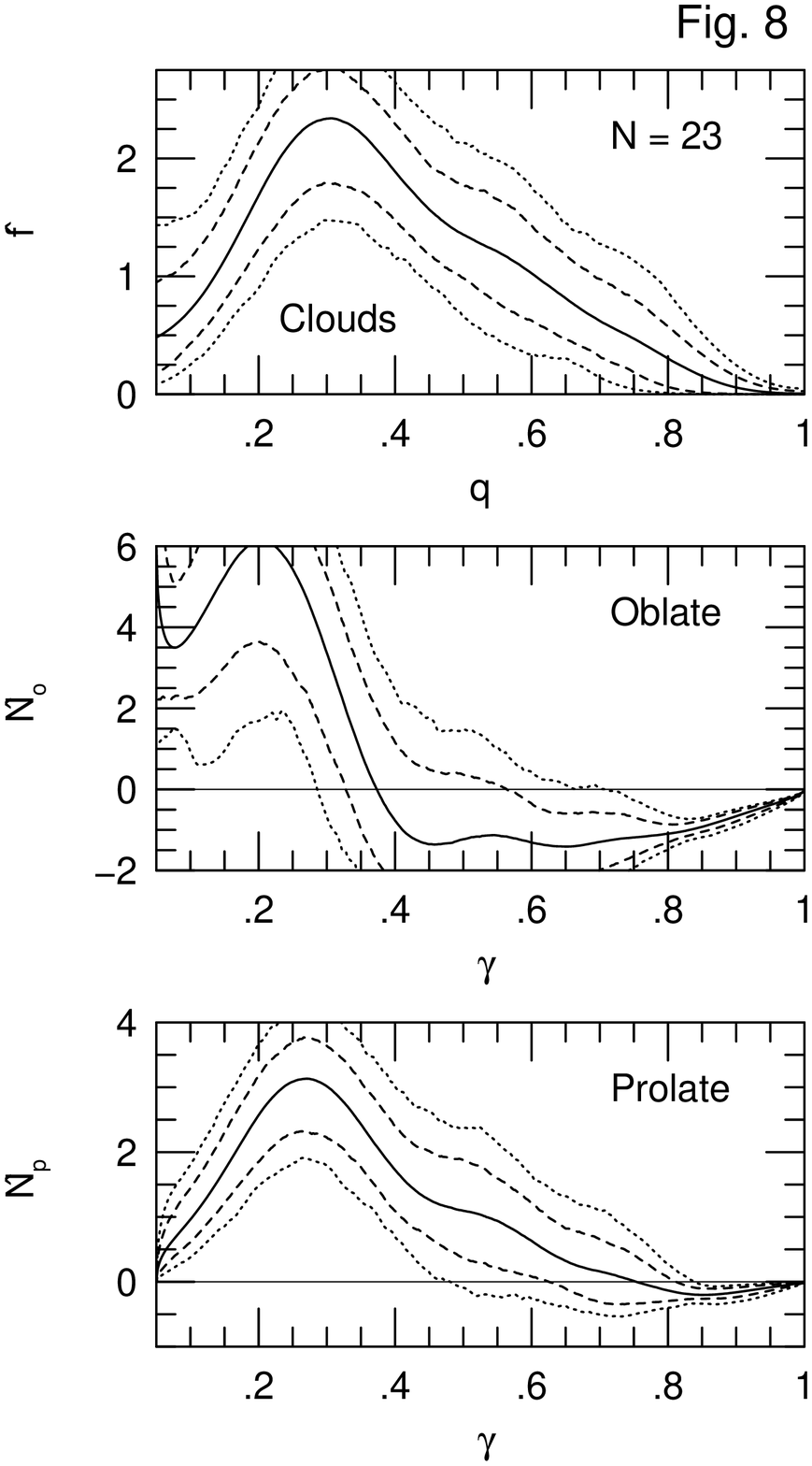}
\end{figure}

\end{document}